\documentclass[aps,superscriptaddress,prl,twocolumn]{revtex4-1}
\usepackage{amsmath}
\usepackage{epsfig}
\usepackage{graphicx}
\usepackage{multirow}
\usepackage{array}
\usepackage{tabularx}
\usepackage{inputenc}
\usepackage{bm}
\usepackage{color}
\usepackage{gensymb}

\newcommand{\ndzr}{Nd$_2$Zr$_2$O$_7$}
\newcommand{\erti}{Er$_2$Ti$_2$O$_7$}
\newcommand{\ersn}{Er$_2$Sn$_2$O$_7$}
\newcommand{\ybti}{Yb$_2$Ti$_2$O$_7$}
\newcommand{\tbti}{Tb$_2$Ti$_2$O$_7$}
\newcommand{\dyti}{Dy$_2$Ti$_2$O$_7$}
\newcommand{\ndir}{Nd$_2$Ir$_2$O$_7$}

\newcommand{\nd}{Nd$^{3+}$}

\newcommand{\er}{Er$^{3+}$}

\begin{document}
\author{E. Lhotel}
\email[]{elsa.lhotel@neel.cnrs.fr}
\affiliation{Institut N\'eel, CNRS and Universit\'e Joseph Fourier, BP 166, 38042 Grenoble Cedex 9, France}
\author{S. Petit}
\email[]{sylvain.petit@cea.fr}
\affiliation{Laboratoire L\'eon Brillouin, CEA-CNRS UMR 12, CE-Saclay, F-91191 Gif-sur-Yvette, France}
\author{S. Guitteny}
\affiliation{Laboratoire L\'eon Brillouin, CEA-CNRS UMR 12, CE-Saclay, F-91191 Gif-sur-Yvette, France}
\author{O. Florea}
\affiliation{Institut N\'eel, CNRS and Universit\'e Joseph Fourier, BP 166, 38042 Grenoble Cedex 9, France}
\author{M. Ciomaga Hatnean}
\affiliation{Department of Physics, University of Warwick, Coventry, CV4 7AL, United Kingdom}
\author{C. Colin}
\affiliation{Institut N\'eel, CNRS and Universit\'e Joseph Fourier, BP 166, 38042 Grenoble Cedex 9, France}
\author{E. Ressouche}
\affiliation{INAC SPSMS, CEA and Universit\'e Joseph Fourier, F-38000 Grenoble, France}
\author{M. R. Lees}
\affiliation{Department of Physics, University of Warwick, Coventry, CV4 7AL, United Kingdom}
\author{G. Balakrishnan}
\affiliation{Department of Physics, University of Warwick, Coventry, CV4 7AL, United Kingdom}

\title{Fluctuations and all-in$-$all-out ordering in dipole-octupole \ndzr}

\begin{abstract}
We present an experimental study of the pyrochlore coumpound \ndzr\ by means of neutron scattering and magnetization measurements down to 90 mK. The Nd$^{3+}$ magnetic moments exhibit a strong local $\langle$111$\rangle$ Ising anisotropy together with a dipolar-octupolar nature, different from the standard Kramers-doublet studied so far. We show that, despite the positive Curie-Weiss temperature, \ndzr\ undergoes a transition around 285 mK towards an all-in$-$all-out antiferromagnetic state. We establish the $(H,T)$ phase diagram in the three directions of the applied field and reveal a metamagnetic transition around 0.1 T. The strongly reduced ordered magnetic moment as well as the unexpected shape of the magnetization curves demonstrate that \ndzr\, is not a standard Ising antiferromagnet. We propose that the peculiar nature of the Nd doublet combined with competing interactions explain these findings.
\end{abstract}

\maketitle
%

Understanding, characterizing and classifying novel states of matter is one of the main issues of research in frustrated systems \cite{Lacroix,Gardner10}. Much of the recent activity in this field has focused on 4{\it f} pyrochlores. Here, the frustration arises from the lattice, built from tetrahedra connected by their corners. The situation where the magnetic moments on these vertices are coupled via a ferromagnetic nearest neighbors interaction and constrained by the crystal electric field (CEF) anisotropy to lie along local $\langle$111$\rangle$ axes, has retained a lot of attention, as it leads classically to a macroscopically degenerated ground state, where each tetrahedron has two spins pointing in and two spins pointing out, called ``spin ice" \cite{harris97,Ramirez99}. 
The quest for the quantum variant of spin ice (QSI) has attracted further strong interest. A wealth of fascinating phenomena, like the emergence of gauge fields along with unconventional excitations are expected \cite{gingras1}. Those systems yet remain to be explored from an experimental point of view. In this search, systems with positive Curie-Weiss temperature $\theta_{\rm CW}$ indicative of resulting ferromagnetic interactions along with strong effective Ising-like anisotropy are {\it a priori} possible candidates. In this letter, we focus on \ndzr\, which falls into this class, as shown by magnetization \cite{blote69,matsuhira,monica} and CEF studies \cite{monica,lutique}. 

In addition, \ndzr\, stands out through an interesting property. In many of the 4{\it f} pyrochlores, the crystal field and spin orbit coupling lead to Kramers ground doublets, well separated from the excited states, so that the low energy physics can 
be described by a pseudo spin 1/2 \cite{rossprx}. In most of them, this spin 1/2 transforms as a magnetic dipole under space group operations, just like a true spin 1/2 moment. This is for instance the case in \erti\, \cite{savary,erti}, \ersn\, \cite{ersn} or \ybti\, \cite{rossprx}. 
However, there exists Kramers doublets called ``dipolar-octupolar" (DO) doublets, where, in a pseudo spin 1/2 representation, two components transform like a magnetic dipole, while the other transforms like the component of the magnetic octupole tensor \cite{abragam}.
This is the case in \ndzr\, \cite{monica} (but also in \ndir\, \cite{watahiki} and \dyti\, \cite{bertin}). The theoretical phase diagram of such systems is highly interesting \cite{hermele}: besides ordered phases like the all-in$-$all-out structure (AIAO) where dipoles point along the local axes toward or away from the tetrahedron centers, an antiferro-octupolar phase appears along with QSI phases.

In this letter, we first confirm the strong Ising character of the \nd\ ion in \ndzr\, and the DO nature of its ground state doublet. 
Combining neutron and magnetization measurements, we show that despite a positive $\theta_{\rm CW}$, \ndzr\, orders in an antiferromagnetic AIAO structure below $T_{\mathrm N}=285$~mK with a strongly reduced magnetic moment. We then establish the $(H,T)$ phase diagram: a metamagnetic-like transition occurs around 0.1 T, and a conventional field induced ordered state is recovered at about 0.2 T. 
This can be reproduced by mean field calculations with an effective antiferromagnetic coupling that however contradicts $\theta_{\rm CW} > 0$. The reduced ordered magnetic moment and the peculiar shape of the magnetization versus field curves can be accounted for by considering local fluctuations induced by octupolar correlations. We propose that, despite the presence of ferromagnetic interactions inferred from positive $\theta_{\rm CW}$ that should stabilize the spin ice state, the peculiar nature of the ground state doublet combined with competing interactions induce the AIAO structure.

Magnetization and ac susceptibility measurements were performed in two SQUID magnetometers equipped with dilution refrigerators developed at the Institut N\'eel \cite{Paulsen01}. The experiments were carried out on a powder sample and a parallelepiped single crystal of \ndzr\, grown by the floating zone technique ($2.35 \times 1.72 \times 1.36$ mm$^3$) \cite{monica1} down to 90 mK. The magnetic field was applied along the three cubic high symmetry directions [100], [110] and [111]. Powder neutron diffraction experiments have been conducted on D1B (CNRS CRG - ILL) using a wavelength $\lambda=2.52$ \AA. We repeated the same experiment on a better quality sample (the one used for macroscopic measurements) on 
4F2 (LLB), using $\lambda=2.36$ \AA. 
Single crystal diffraction experiments were carried out on D23 (CEA CRG - ILL) using $\lambda=1.28$ and 2.4 \AA. 
Finally, CEF excitations were measured on the high quality powder sample by means of inelastic neutron scattering conducted on the thermal triple axis spectrometer 2T (LLB). 

\begin{figure}[h!]
\includegraphics[width=8cm]{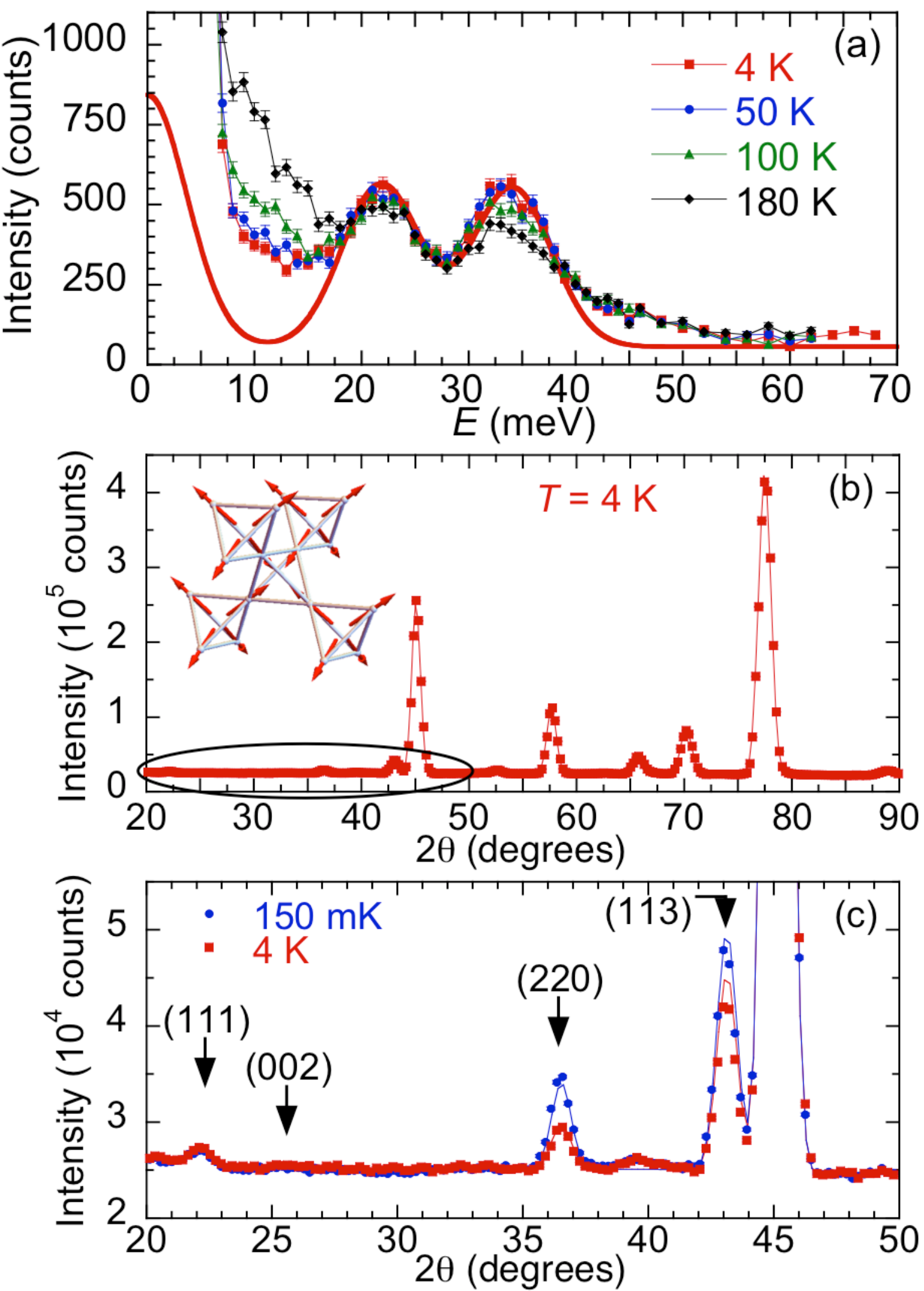}
\caption{\label{fig_neutrons} (a) Inelastic neutron scattering spectra taken at several temperatures.
The solid line shows the fit with the coefficients of Table \ref{DO}.
(b) Powder diffractogram at 4 K. The line is a fit to the $Fd\bar{3}m$ structure. Inset: Scheme of the AIAO state. 
(c) Zoom on the low angle part at 150 mK and 4 K showing the magnetic contribution. The lines are combined refinements of the crystallographic and magnetic structure. 
}
\end{figure}

First, we focus on the DO nature of the ground doublet in \ndzr\, and study the CEF excitations. 
As shown in Figure \ref{fig_neutrons}(a), the spin-spin correlation function $S(Q,\omega)$ exhibits two clear peaks at 21 and 33 meV. 
The intensity of both excitations tends to slightly decrease as the temperature increases and remains roughtly constant at larger wavevector $Q$. 
These two characteristics are evidence for the magnetic origin of both modes, that are thus attributed to two CEF transitions. 
A better energy resolution reveals the presence of an additional single mode at 10~meV. Its temperature and $Q$ evolution, strong increase with increasing temperature or increasing Q, are typical of a phonon mode. The energy of these two CEF modes match the theoretical analysis based on macroscopic measurements, that takes into account the mixing of the $^4I_{9/2}$ and $^4I_{11/2}$ multiplets of the $4f^3$ \nd\, ions electronic configuration \cite{monica}. Note that the theory predicts two distinct modes around 32 meV which could not be resolved in the experiment \cite{monica2}. 
To explicitely show the DO character of the doublet, and determine wavefunctions easier to handle, we consider the CEF Hamiltonian within the ground multiplet $^4I_{9/2}$ only:   
${\cal H}_{\mbox{CEF}}= \sum_{m,n} B_{nm} O_{nm}$ where the $O_{nm}$ are the Stevens operators ($J=9/2$, $g_J=8/11$) \cite{wybourne}. The $B_{nm}$ coefficients are refined by fitting the neutron data (energy levels and intensities) along with the value of the local magnetic moment at low temperature 2.3 $\mu_B$ (see Figure \ref{fig_neutrons}(a) and Table \ref{DO}). This procedure 
gives modes at 22, 33.5, 34.5 and 109 meV yielding a picture of very strong Ising \nd\, moments 
with $g_{\perp} \equiv 0$ and $g_ {//}=4.3$ in agreement with Ref. \cite{monica}. The coefficients of the ground doublet wave-functions $|\sigma=\uparrow,\downarrow \rangle$ (see Table \ref{DO}) are similar to the ones found in \ndir\, \cite{watahiki} and typical of a DO doublet \cite{abragam}. 

\begin{table}
\begin{tabularx}{\linewidth}{*{8}{>{\centering\arraybackslash}X}}
\hline \hline
$B_{20}$& $B_{40}$ &$ B_{43}$ & $B_{60}$ &$ B_{63}$ & $B_{66}$ & $g_{\parallel} $& $g_{\perp} $ \\ \hline
-190&5910&110&2810&10&-890&4.3&0 \\ \hline \hline
&&&&&&&
\end{tabularx}
\begin{tabularx}{\linewidth}{p{0.6cm}*{8}{c}>{\centering\arraybackslash}X} \hline \hline
 $J_z$& $-9/2$ & $-7/2$ & $-5/2$ & $-3/2$ & $\pm1/2$ & $3/2$ &$ 5/2$ &$ 7/2$ & $9/2$ \\ \hline
$| \uparrow\rangle$   & -0.878 & 0 & 0 & $-0.05$ & 0 & 0.476 & 0 & 0 & $0.009$ \\
$| \downarrow\rangle$ &-0.009 & 0 & 0 & 0.476 & 0 & 0.05 & 0 & 0 & -0.878 \\
\hline \hline
\end{tabularx}
\caption{Stevens coefficients (in K), $g$-factors and ground state wave-functions of \nd\ in \ndzr\ from inelastic neutron scattering analysis. }
\label{DO}
\end{table}

\begin{figure}
\includegraphics[width=8cm]{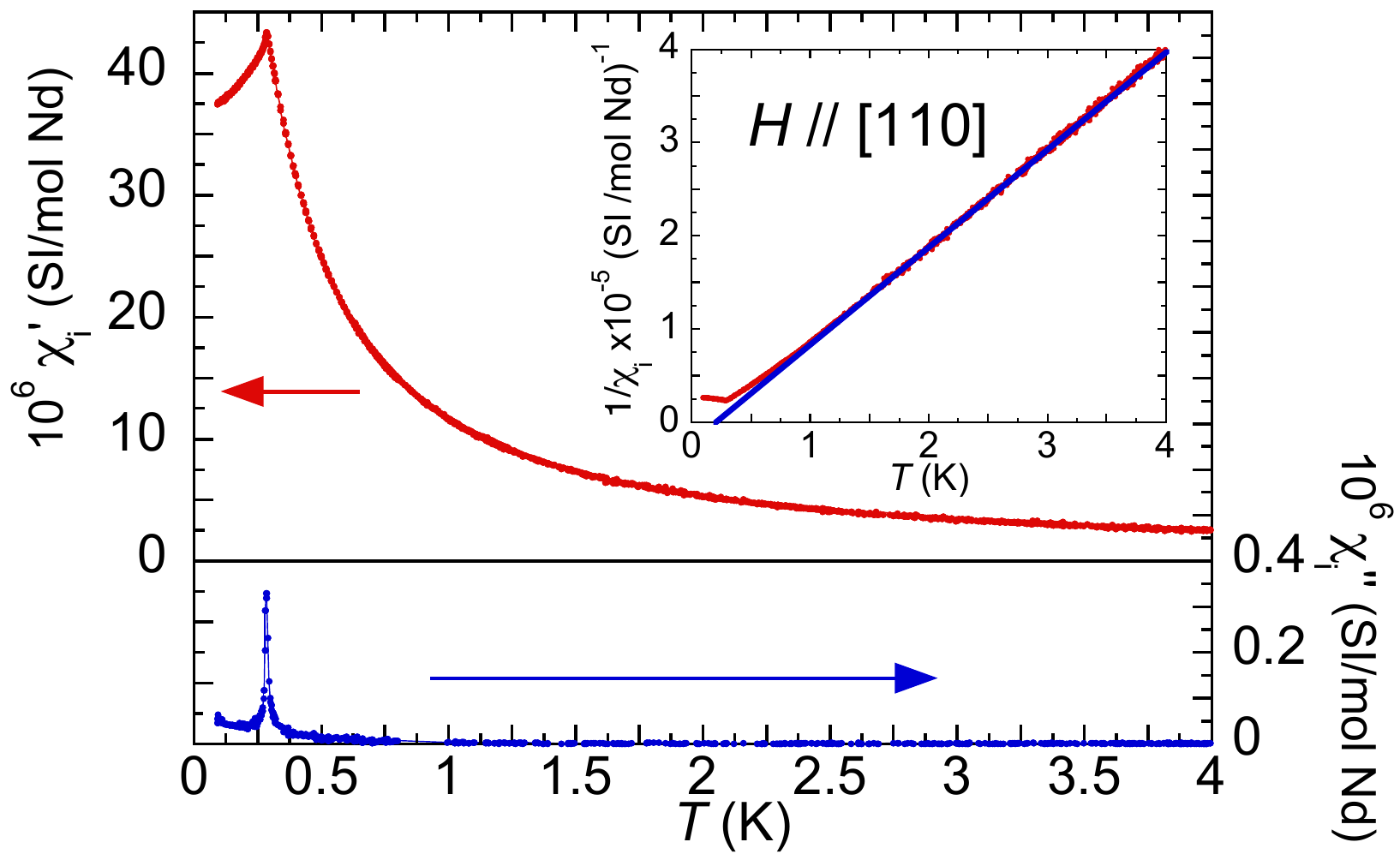}
\caption{\label{fig_XT_HT} ac susceptibility $\chi_i'$ and $\chi_i''$ vs temperature $T$ along [110] with $H_{\rm ac}=2.7$ Oe and $f=5.7$ Hz, corrected for demagnetization effects. Inset: $1/\chi'_i$ vs $T$. The line is a fit between 1 and 4 K to the Curie-Weiss law $1/\chi=2.05 \times 10^4 -1.05\times 10^5~T$, giving $\theta_{CW}=195~(\pm 15)$~mK and an effective moment $\mu_{\rm eff}=2.45~(\pm 0.02)~\mu_B$.}
\end{figure}

Our magnetization and ac susceptibility measurements reveal a transition at low temperature, as expected from 
specific heat measurements \cite{blote69}. 
The susceptibility exhibits a peak both in the real and imaginary parts at a temperature of 410 mK for the powder sample, consistent with ref. \cite{blote69} (not shown), and at a lower temperature of 285 mK for the single crystal (see Figure \ref{fig_XT_HT}).
Despite $\theta_{\rm CW} > 0$ 
(see inset of Figure \ref{fig_XT_HT}), the shape of the $\chi'$ peak, as well as its amplitude are characteristic of an antiferromagnetic transition. The position of the peak does not depend on frequency (in the measured range $f=0.11-570$~Hz). However, when the frequency increases, the peak of 
$\chi''$ lies on top of a background below 4 K which increases 
with decreasing temperature, similarly to what was observed in \tbti\, \cite{Lhotel12} (not shown). This observation might be the signature of slow fluctuations
, not affected by the antiferromagnetic transition. 

This antiferromagnetic transition is further confirmed by neutron diffraction measurements. At 4 K (see Figure \ref{fig_neutrons}(b)), the \textsc{fullprof} refinement \cite{fullprof} in the $Fd\bar{3}m$ space group yields a lattice parameter $a=10.66$ \AA\ and the O2 oxygen position $x=0.335$ (with an RF factor of 2). 
A magnetic contribution rises below $T_N$ and appears on the nuclear Bragg peaks, indicating that the propagation vector of the magnetic structure is ${\bf k}=(0,0,0)$ (see Figure \ref{fig_neutrons}(c)). The symmetry analysis shows that a number of magnetic configurations, corresponding to the different irreducible representations $\Gamma_{3,5,7,9}$ are in principle possible candidates \cite{poole}. $\Gamma_3$ corresponds to the AIAO structure (see inset of Figure \ref{fig_neutrons}(b)) predicted for multi-axis Ising pyrochlore antiferromagnets \cite{Bramwell98}, $\Gamma_5$ to the structure of the antiferromagnetic XY pyrochlore magnet Er$_2$Ti$_2$O$_7$ \cite{Champion03}, $\Gamma_7$ to the structure of dipolar pyrochlore magnets such as Gd$_2$Sn$_2$O$_7$ \cite{Palmer00, Wills06}, and $\Gamma_9$ to the splayed ferromagnetic Yb$_2$Sn$_2$O$_7$ structure \cite{Yaouanc13, Wills06}. 
Our neutron data are clearly compatible with $\Gamma_3$ since it is the sole configuration that leads to a vanishing contribution of the (111) and (002) (plus symmetry related) Bragg peaks, as observed in the experiment. The \textsc{fullprof} refinement 
of the diffractogram yields a weak magnetic moment $m=0.8\pm 0.05~\mu_B$ indicating large fluctuations even at temperatures of about $T_N/4$. Diffraction measurements performed on a single crystal confirm this analysis and lead to the same conclusion. Note that this AIAO ordering, although classically expected in Ising pyrochlore antiferromagnets, has rarely been reported. It has been observed in compounds such as FeF$_3$ \cite{Ferey}, Cd$_2$Os$_2$Sb$_7$ \cite{Yamaura12} or Na$_3$Co(CO$_3$)$_2$Cl \cite{Fu13}, but in these cases, it does not result from Ising anisotropy. The ordering of the rare-earth moments in pyrochlore iridates \cite{Tomiyasu12, Lefrancois15} may be placed in this category, but this ordering is induced by the AIAO state of the iridium magnetic moments and, {\it a priori}, not to the magnetic interactions between rare-earth moments.  In that sense, the observation of the AIAO state itself makes \ndzr\ a peculiar case in the insulating pyrochlore compounds.

\begin{figure}
\includegraphics[width=7.5cm]{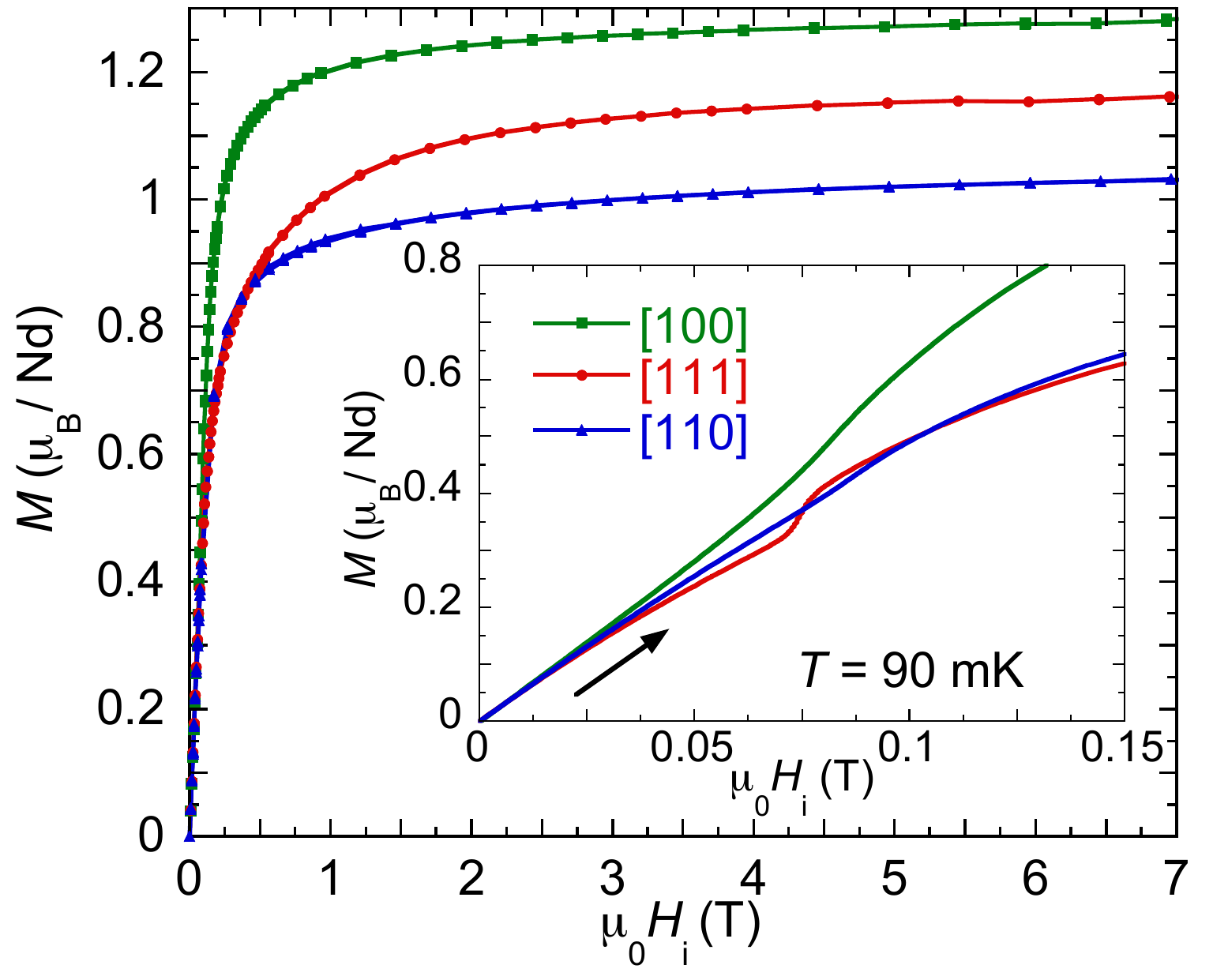}
\caption{\label{fig_MH} Magnetization $M$ vs internal field $H_i$ measured along the three main directions at 90 mK. Inset: Detailed view of the low field data. The field was swept up from negative to positive values.}
\end{figure}

\begin{figure}
\includegraphics[width=8cm]{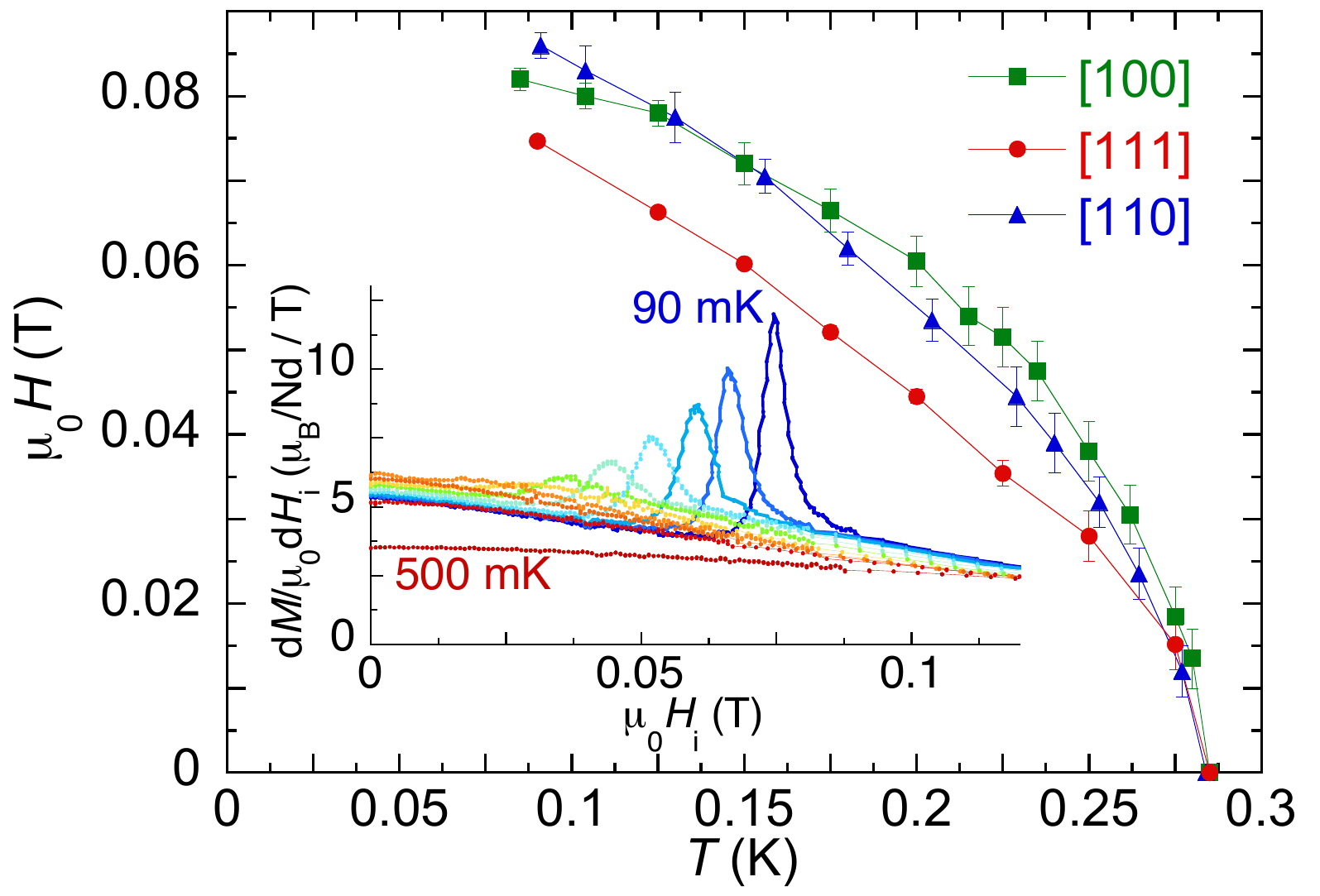}
\caption{Phase diagram $(H,T)$ obtained from the maximum of $dM/dH_i$ vs $H_i$. Inset: $dM/dH_i$ vs $H_i$ in the [111] direction. }
\label{phasdiag}
\end{figure}

Aiming at a full characterization of this system, we proceed with the determination of the $(H,T)$ phase diagram by performing $M$ vs $H$ measurements down to 90~mK. As previously reported \cite{monica}, the saturated magnetization depends on the field direction (see Figure \ref{fig_MH}). The obtained values are, within 10\%, the ones expected for magnetic moments with a multi-axis Ising anisotropy \cite{harris98} and correspond to a total magnetic moment $m \approx 2.3~\mu_B$, consistent with the effective moment of the Curie-Weiss law (see inset of Figure \ref{fig_XT_HT}). This shows that, although the ordered moment is reduced in the AIAO phase, the total magnetic moment is recovered in high magnetic fields, and remains with a strong Ising character. At low field, the slope of the magnetization curves is not zero (see inset of Figure \ref{fig_MH}), contrary to what is expected for a strongly anisotropic antiferromagnet. This confirms that a part of the magnetic moment is still fluctuating at low temperature. An inflexion point is present in these curves around 0.1 T, which can be attributed to a metamagnetic process towards the field induced ordered phase. It is observed in the three directions of the applied field, but is more pronounced along the [111] direction. The temperature dependence of this metamagnetic process can be followed by plotting the derivative $dM/dH_i$ of the magnetization curve (see inset of Figure \ref{phasdiag}). Assuming that the maximum of $dM/dH_i$ corresponds to the metamagnetic field value, we plot the phase diagram as a function of temperature (see Figure \ref{phasdiag}). As expected the $dM/dH_i$ peak disappears at the transition temperature. The obtained phase diagram is almost isotropic, and the data can be fitted to a $T-T_N$ law which gives an exponent slightly larger than 1/2. 

\begin{figure}
\includegraphics[width=8.5cm]{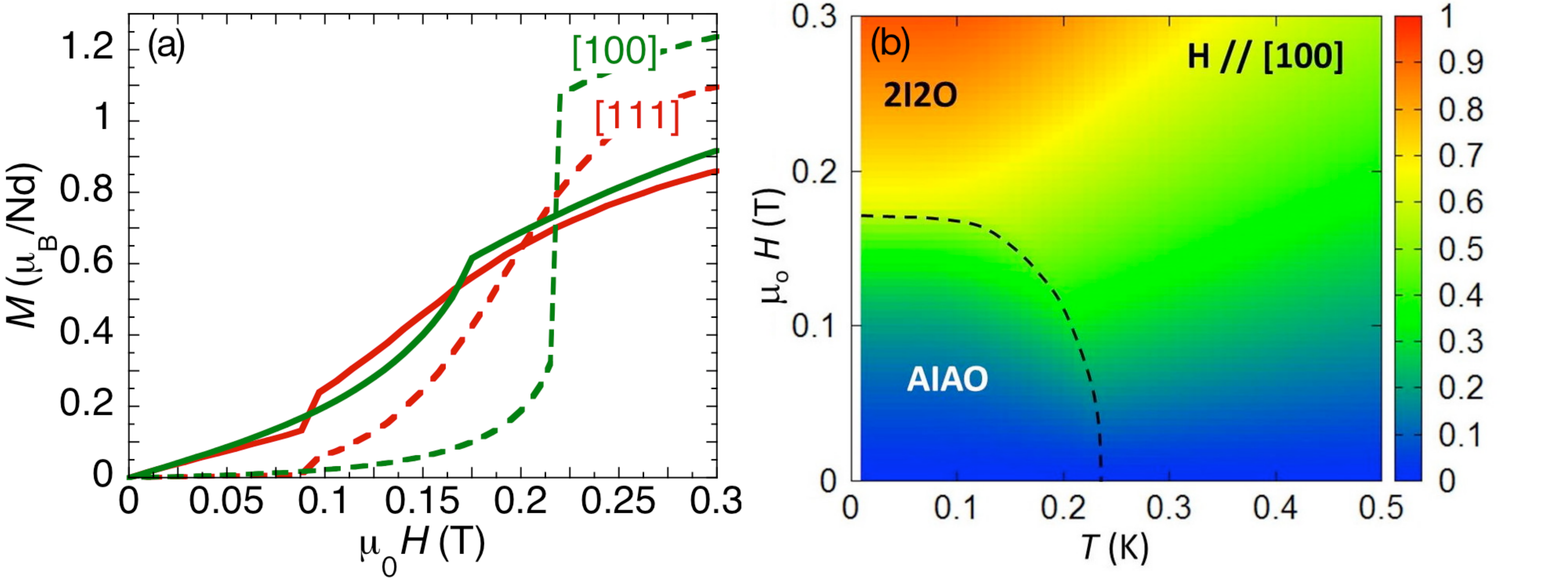}
\caption{(a) Calculated $M$ vs $H$ at $T=100$ mK with ${\sf J}^z \sim -0.005$ K for the field applied along the [100] (green) and [111] (red) directions. Full (resp. dashed) lines are calculated with $U=0.2$ K (resp. $U=0$). (b) Calculated Phase diagram with $H \parallel [100]$ superimposed with $M$. ${\sf J}^z \sim -0.005$ K and $U=0.2$ K. The dashed line shows the transition.}
\label{fig_calc}
\end{figure}

We now compare these experimental results with mean-field calculations. To this end, we consider a model taking into account ${\cal H}_{\mbox{CEF}}$ and a unique coupling ${\sf J}^z$ acting between nearest neighbor local $z$ components of the Nd\, moments. 
An antiferromagnetic ${\sf J}^z \sim -0.005$ K allows to reproduce the N\'eel temperature of about 300 mK and the AIAO structure. 
This leads to a $(H,T)$ phase diagram 
 similar to the experimental one, with a metamagnetic-like transition from the AIAO state to the field induced ordered state (3-in$-$1-out or 2-in$-$2-out state depending on the direction of the field). 
Nevertheless, this mean-field model fails to reproduce the positive $\theta_{\rm CW}$ as well as the small amplitude of the ordered moment. In addition, the calculated field induced transition is abrupt, corresponding to spin-flips in the whole magnetic structure, in contrast with the smooth measured curves (see dashed lines in Figure \ref{fig_calc}(a)). 

These results support the existence of strong fluctuations 
which prevent a total ordering of the moment. Further, from the coefficients of the DO wavefunction, $\langle \uparrow | J | \downarrow \rangle \equiv 0$, so the dipolar exchange alone cannot induce any on-site fluctuation from one element of the doublet to the other. This suggests that another coupling is at play. Considering the DO nature of the ground wavefunction, such fluctuations arise by virtue of the octupolar operator ${\cal T}=i(J^+J^+J^+-J^-J^-J^-)$ \cite{hermele}. Indeed, using Table \ref{DO}, one has $\langle \uparrow | {\cal T} | \downarrow \rangle \ne 0$. As a result, an octupolar field induces a mixing term in the Hamiltonian ${\cal V} = \sum_{\sigma=\uparrow,\downarrow}| \sigma \rangle U \langle -\sigma |$ which tends to bind the two elements of the doublet to form new states with reduced moments. In this picture, applying a magnetic field competes with ${\cal V}$ and progressively unbinds $| \uparrow \rangle$ and $| \downarrow \rangle$. The effect of such term is consistent with the experimental observations, since, as shown in Fig. \ref{fig_calc}, it does reduce the ordered moment, and makes the magnetization curves and field induced transition smoother. However, the calculation of the susceptibility with this value of ${\sf J}^z$ and taking into account ${\cal V}$ still cannot reproduce the sign of $\theta_{\rm CW}$. 

This discrepancy might be understood in the context of the pseudo spin 1/2 theory developed recently for DO doublets \cite{hermele}. It is shown that despite a positive ${\sf J}^z$ (corresponding to a ferromagnetic-like interaction in that local frame convention), the antiferromagnetic AIAO ground state is stabilized by a negative transverse term ${\sf J}^x$ in a large region of the theoretical phase diagram. But due to the peculiar nature of the ground state doublet, only ${\sf J}^z$ contributes to the Curie-Weiss temperature, leading to a positive $\theta_{\rm CW}$. 
Further analysis of spin waves would help to determine these coupling constants and serve as a test of this hypothesis. 

It is worth noting that this contradiction might also be removed in the context of the recently proposed fragmentation theory \cite{holdsworth}. In this model, for a certain range of microscopic parameters, the classical spin-ice phase can fragment into an AIAO antiferromagnetic phase with a reduced moment, and a Coulomb phase characteristic of the spin-ice correlations. Such a scenario could be confirmed by diffuse scattering measurements, showing the coexistence of Bragg peaks and pinch points. 

In summary, our inelastic neutron scattering results confirm the Ising and DO nature of the ground state doublet of the \nd\ ion. Despite a Curie-Weiss temperature indicative of ferromagnetic interactions, the existence of an AIAO ordering is established. Strong fluctuations, unexpected in an Ising magnet, reduce however the magnetic moment. We also provide evidence for a metamagnetic transition towards a field-induced ordered state below the N\'eel temperature. More generally, this study points out how quantum fluctuations are reintroduced in Ising systems by multipolar correlations. In \ndzr, the DO nature of the doublet naturally calls for octupolar correlations. Further theoretical and experimental work is needed especially to determine the nature of magnetic excitations stemming from this peculiar ground state.




\begin{thebibliography}{99}
\bibitem{monica2} An additional mode is predicted at 100 meV, but is not visible in the accessible energy range. Note that the observed CEF levels do not match those proposed in \cite{lutique}.
\end{thebibliography}
\end{document}